\documentstyle[psfig]{article}
\newcommand{\gtequiv}{\lower2pt\hbox{$\:\stackrel{>}{
\scriptstyle\sim}\:$}}
\newcommand{\ltequiv}{\lower2pt\hbox{$\:\stackrel{<}{
\scriptstyle\sim}\:$}}
\newcommand{\case}[2]{\mbox{${\textstyle\frac{#1}{#2}}$}}

\newcommand{\one}{\mbox{\large\bf 1}}

\begin{document}

\noindent
{\small Published in: {\em Transport Phenomena in Mesoscopic Systems},
edited by H. Fukuyama and T. Ando (Springer, Berlin, 1992).
}\bigskip\\
{\Large\bf Three ``universal'' mesoscopic Josephson effects
}\bigskip\\ C. W. J. Beenakker
\bigskip\\ Instituut-Lorentz, University of Leiden\\
P.O. Box 9506, 2300 RA Leiden, The Netherlands
\bigskip\\ {\bf Abstract.}
A recent theory is reviewed for the sample-to-sample fluctuations
in the critical current of a Josephson junction consisting
of a disordered point contact or microbridge.
The theory is based on a relation between the supercurrent and
the scattering matrix in the normal state. The root-mean-square
amplitude ${\rm rms}\, I_{\rm c}$ of the critical current
$I_{\rm c}$ at zero temperature is given by
${\rm rms}\, I_{\rm c} \simeq e\Delta_{0}/\hbar$,
up to a numerical coefficient of order unity ($\Delta_{0}$ is the energy gap).
This is the superconducting analogue of
``Universal Conductance Fluctuations'' in the normal state.
The theory can also be applied to a ballistic point contact, where it yields
the analogue of the quantized conductance, and to a quantum dot, where it
describes supercurrent resonances. All three phenomena provide a measurement of
the supercurrent unit $e\Delta_{0}/\hbar$, and are ``universal'' through the
absence of a dependence on junction parameters.

\section{Introduction}

Nanostructures combining semiconducting and superconducting elements
form a new class of systems in which to search for mesoscopic
phenomena. The correspondence between transport of normal electrons
and transport of Bogoliubov quasiparticles (being the elementary
excitations of the superconductor) serves as a
useful guide in the search. The appearance of a new length scale
--- the superconducting coherence length $\xi$ ---
complicates the correspondence in an interesting way, by
introducing two qualitatively different regimes:
The {\em short-junction} regime (junction length $L\ll\xi$)
and the {\em long-junction} regime ($L\gg\xi$).
These two regimes appear over and above the transport regimes
which depend on the relative magnitude of junction length
and mean free path $l$: The {\em ballistic\/} regime ($l\gg L$) and
the {\em diffusive\/} regime ($l\ll L$). A third transport regime in
the normal state, that of {\em resonant tunneling}, is also subdivided into two
new regimes, distinguished by the relative magnitude of $\xi$ and the
characteristic length $v_{\rm F}\tau_{\rm res}$, with $\tau_{\rm res}$ the
lifetime of the resonant state in the junction and $v_{\rm F}$ the Fermi
velocity.

Our own interest in this field has focused on a set of three phenomena which
provide, through the Josephson effect, a measurement of the supercurrent unit
$e\Delta_{0}/\hbar$ ($\Delta_{0}$ being the superconducting energy gap). This
unit of current plays the role of the conductance quantum $e^{2}/h$ in the
normal state. The three phenomena are:
\begin{enumerate}
\item Discretization of the critical current
of a ballistic point contact \cite{Bee91}.
\item Resonant Josephson current through a quantum dot \cite{Gla89,SQUID91}.
\item Mesoscopic supercurrent fluctuations in a diffusive point contact
\cite{Bee91a}.
\end{enumerate}
These effects belong, respectively, to the ballistic, resonant-tunneling, and
diffusive transport regimes, and within each regime to the short-junction
limit.

Effect number 1 is the analogue of the quantized conductance of a quantum point
contact \cite{Wee88}--\cite{Eer91}: The critical current $I_{\rm c}$ of a short
and narrow constriction in a superconductor increases stepwise as a function of
the constriction width, with step height $e\Delta_{0}/\hbar$ independent of the
properties of the junction \cite{Bee91},
\begin{eqnarray}
I_{\rm c}=N\frac{e\Delta_{0}}{\hbar}.\label{effect1}
\end{eqnarray}
The integer $N$ is the number of transverse modes at the Fermi level which can
propagate through the constriction. The short-junction limit $L\ll\xi_{0}$
(where $\xi_{0} =\hbar v_{\rm F}/\pi\Delta_{0}$ is the ballistic coherence
length) is essential: Furusaki et al.\ \cite{Fur91} studied quantum size
effects on the critical current in the long-junction limit and found a
geometry-dependent behavior instead of Eq.\ (\ref{effect1}).

Effect number 2 is described by the formula \cite{SQUID91}
\begin{eqnarray}
I_{\rm c}^{\rm
res}=\frac{e}{\hbar}\,\frac{\Delta_{0}\Gamma_{0}}{\Delta_{0}+\Gamma_{0}}
\label{effect2a}
\end{eqnarray}
for the critical current through a bound state at the Fermi level which is
coupled with equal tunnel rate $\Gamma_{0}/\hbar$ to two bulk superconductors.
Since $\tau_{\rm res}=\hbar/2\Gamma_{0}$, the ``short-junction'' criterion
$\xi_{0}\gg v_{\rm F}\tau_{\rm res}$ is equivalent to
$\Gamma_{0}\gg\Delta_{0}$. If $\Gamma_{0}\gg\Delta_{0}$, the critical current
on resonance becomes \cite{Gla89,SQUID91}
\begin{eqnarray}
I_{\rm c}^{\rm res}=\frac{e\Delta_{0}}{\hbar}, \label{effect2b}
\end{eqnarray}
which no longer depends on the tunnel rate.

Effect number 3 is the analogue of ``Universal Conductance Fluctuations''
in disordered normal metals \cite{Alt85}--\cite{Alt91}. The sample-to-sample
fluctuations in the critical current of a disordered point contact have
root-mean-square value \cite{Bee91a}
\begin{eqnarray}
{\rm rms}\, I_{\rm c}\simeq\frac{e\Delta_{0}}{\hbar},\label{effect3}
\end{eqnarray}
up to a numerical coefficient of order unity. These mesoscopic fluctuations are
{\em universal}\/ in the sense that they do not depend on the size of the
junction or on the degree of disorder, as long as the criteria $l\ll L\ll Nl$
and $L\ll\xi$ for the diffusive, short-junction regime are satisfied. (Here
$\xi =(\xi_{0}l)^{1/2}$ is the diffusive coherence length.) The short-junction
limit $L\ll\xi$ is again essential for universality. The opposite long-junction
limit was considered previously by Al'tshuler and Spivak \cite{Alt87}, who
calculated the supercurrent fluctuations of an SNS junction in the limit
$L\gg\xi$ (S=superconductor, N=normal metal, $L$=separation of the SN
interfaces). Their result ${\rm rms}\, I_{\rm c}\simeq ev_{\rm F}l/L^{2}$
depends on both $l$ and $L$ and is therefore {\em not\/} universal in the sense
of UCF.

We have summarized the three universal short-junction Josephson effects in
Table 1, together with their non-universal long-junction counterparts. We have
introduced the diffusion constant $D\simeq v_{\rm F}l$ and the correlation
energy (or Thouless energy) $E_{\rm c}\simeq\hbar/\tau_{\rm dwell}$, where
$\tau_{\rm dwell}$ is the dwell time in the junction. The distinction
``short-junction'' versus ``long-junction'' may alternatively be given in terms
of the relative magnitude of $E_{\rm c}$ and $\Delta_{0}$: In the
short-junction regime $\Delta_{0}\ll E_{\rm c}$, while in the long-junction
regime $\Delta_{0}\gg E_{\rm c}$. The universality is lost in the long-junction
regime because $E_{\rm c}$ (which depends on junction parameters) becomes the
characteristic energy for the Josephson effect instead of $\Delta_{0}$. All
these results hold in the zero-temperature limit, or more precisely for $k_{\rm
B}T\ll{\rm min}\, (\Delta_{0},E_{\rm c})$.

\begin{table}[tb]
\centering
\begin{tabular}{l|cc|c}
\multicolumn{4}{c}{Mesoscopic Josephson Effects}\\ \hline
& short-junction & long-junction & $E_{\rm c}$\\ \hline
ballistic & $I_{\rm c}=Ne\Delta_{0}/\hbar$ & geometry dep. & $\hbar v_{\rm
F}/L$\\
res.tunneling & $I_{\rm c}^{\rm res}=e\Delta_{0}/\hbar$ & $I_{\rm c}^{\rm
res}=e\Gamma_{0}/\hbar$ & $\Gamma_{0}$ \\
diffusive & ${\rm rms}\, I_{\rm c}\simeq e\Delta_{0}/\hbar$ & ${\rm rms}\,
I_{\rm c}\simeq eD/L^{2}$ & $\hbar D/L^{2}$\\ \hline
\end{tabular}
\caption{Summary of the mesoscopic Josephson effects described in the text, in
the ballistic, resonant-tunneling, and diffusive transport regimes. Universal
(junction-independent) behavior is obtained in the short-junction limit, which
in terms of the correlation energy $E_{\rm c}$ is given by $\Delta_{0}\ll
E_{\rm c}$.}
\end{table}

In the present paper we review a scattering theory for the Josephson effect,
which provides a unified treatment of the three universal (short-junction)
phenomena. In Sec.\ 2 we show, following Ref.\ \cite{SFe91}, how the
supercurrent can be obtained directly from the quasiparticle excitation
spectrum of the Josephson junction. This approach is equivalent to the usual
method which starts from the finite-temperature Greens function, but is more
convenient in the short-junction regime, where --- as we shall see --- the
excitation spectrum has a particularly simple form. In Sec.\ 3 we follow Ref.\
\cite{Bee91a} in relating the excitation spectrum to the normal-state
scattering matrix of the junction. In Sec.\ 4 we take the short-junction limit
and obtain a simple relation between the supercurrent and the eigenvalues of
the transmission matrix product $tt^{\dagger}$. This relation plays the role
for the Josephson effect of the Landauer formula for the conductance
\cite{Eer91}. We emphasize that the transmission matrix $t$ refers to {\em
normal\/} electrons. Other scattering approaches to the Josephson effect
\cite{Fur91a,Wee91} relate the supercurrent to the scattering matrix for
Bogoliubov quasiparticles. The usefulness of the present method is that one can
use directly the many properties of $t$ known from normal-state transport
theory. Indeed, as we shall see in Sec.\ 5, the three universal Josephson
effects follow as special cases of the general scattering formula of Sec.\ 4.

\section{Supercurrent from Excitation Spectrum}

The quasiparticle excitation spectrum of the Josephson junction consists of the
positive eigenvalues of the Bogoliubov-de Gennes (BdG) equation \cite{deG66}.
The BdG equation has the form of
two Schr\"{o}dinger equations for electron
and hole wavefunctions ${\rm u}({\bf r})$ and ${\rm v}({\bf r})$,
coupled by the pair potential $\Delta({\bf r})$:
\begin{eqnarray}
\left(\begin{array}{cc} {\cal H}_{0}&\Delta\\ \Delta^{\ast}&-{\cal
H}_{0} \end{array}\right) \left(\begin{array}{c}{\rm u}\\{\rm
v}\end{array}\right)=\varepsilon \left(\begin{array}{c}{\rm u}\\{\rm
v}\end{array}\right) .\label{BdG1}
\end{eqnarray}
Here ${\cal H}_{0}={\bf
p}^{2}/2m+V({\bf r})-E_{\rm F}$ is the single-electron Hamiltonian, containing
an electrostatic potential $V({\bf r})$ (zero vector potential is assumed). The
excitation energy $\varepsilon$ is measured
relative to the Fermi energy $E_{\rm F}$.

In a uniform system with $\Delta({\bf r})\equiv\Delta_{0}{\rm e}^{{\rm
i}\phi}$, $V({\bf r})\equiv 0$, the
solution of the BdG equation is
\begin{eqnarray}
\varepsilon&=&\left(
(\hbar^{2}k^{2}/2m-E_{\rm F})^{2}+\Delta_{0}^{2}\right) ^{1/2}
,\nonumber\\
{\rm u}({\bf r})&=&{\cal V}^{-1/2}(2\varepsilon)^{-1/2} {\rm e}^{{\rm
i}\phi/2}\left( \varepsilon+ \hbar^{2}k^{2}/2m-E_{\rm F}\right) ^{1/2}
{\rm e}^{{\rm i}{\bf k}\cdot{\bf r}} ,\nonumber\\
{\rm v}({\bf
r})&=&{\cal V}^{-1/2}(2\varepsilon)^{-1/2} {\rm e}^{-{\rm i}\phi/2}\left(
\varepsilon-
\hbar^{2}k^{2}/2m+E_{\rm F}\right) ^{1/2} {\rm e}^{{\rm i}{\bf
k}\cdot{\bf r}}.\label{uniform}
\end{eqnarray}
The eigenfunction is normalized to unit probability in a volume ${\cal V}$,
\begin{eqnarray}
\int_{\cal V} d{\bf r}\, \left( |u|^{2}+|v|^{2}\right) =1.\label{normalize}
\end{eqnarray}
The excitation spectrum
is continuous, with excitation gap $\Delta_{0}$. The eigenfunctions
$({\rm u},{\rm v})$ are plane waves characterized by a wavevector ${\bf
k}$. The coefficients of the plane waves are the two coherence factors
of the BCS theory. This simple excitation spectrum is modified by the presence
of the Josephson junction. The spectrum acquires a discrete part due to the
non-uniformities in $\Delta({\bf r})$ near the junction. The discrete spectrum
corresponds to bound states
in the gap ($0<\varepsilon <\Delta_{0}$), localized within a coherence length
from
the junction. In addition, the continuous spectrum is modified. As we shall
show in Sec.\ 3, in the short-junction limit the supercurrent is entirely
determined by the {\em discrete\/} part of the excitation spectrum.

Consider an SNS junction with normal region at $|x|<L/2$. Let $\Delta ({\bf
r})\rightarrow\Delta_{0}{\rm e}^{\mp{\rm i}\phi/2}$ for
$x\rightarrow\pm\infty$. To determine $\Delta({\bf r})$ near the junction one
has to solve the self-consistency equation \cite{deG66}
\begin{eqnarray}
\Delta({\bf r})=|g({\bf r})|
\sum_{\varepsilon>0}{\rm v}^{\ast}({\bf r}){\rm u}({\bf r})
[1-2f(\varepsilon)] ,\label{selfconsist}
\end{eqnarray}
where the sum is
over all states with positive eigenvalue, and
$f(\varepsilon)=[1+\exp(\varepsilon/k_{\rm B}T)]^{-1}$ is the Fermi function.
The coefficient $g$ is the interaction constant of the BCS theory
of superconductivity. At an SN interface, $g$ drops abruptly (over
atomic distances) to zero. (We assume non-interacting electrons in the normal
region.) Therefore, $\Delta({\bf r})\equiv 0$ for $|x|<L/2$ regardless of Eq.\
(\ref{selfconsist}). At the superconducting side of the SN interface,
$\Delta({\bf r})$ recovers its bulk value $\Delta_{0}$ only at some distance
from the interface (proximity effect). There exists a class of Josephson
junctions where the suppression of $\Delta({\bf r})$ on approaching the SN
interface can be neglected, and one can use the step-function model
\begin{eqnarray}
\Delta({\bf r})=\left\{\begin{array}{ll}
\Delta_{0}{\rm e}^{{\rm i}\phi/2}&\;{\rm if}\;x<-L/2,\\
0 &\;{\rm if}\;|x|<L/2,\\
\Delta_{0}{\rm e}^{-{\rm i}\phi/2}&\;{\rm if}\;x>L/2.
\end{array}\right.\label{Delta0}
\end{eqnarray}
The step-function pair potential is also referred to in the literature as a
``rigid boundary-condition''. Likharev \cite{Lik79} discusses in detail the
conditions under which this model is valid:
\begin{enumerate}
\item If the width $W$ of the junction is small compared to the coherence
length, the non-uniformities in $\Delta({\bf r})$ extend only over a distance
of order $W$ from the junction (because of ``geometrical dilution'' of the
influence of the narrow junction in the wide superconductor). Since
non-uniformities on length scales $\ll\xi$ do not affect the dynamics of the
quasiparticles, these can be neglected and the step-function model holds.
Constrictions in the short-junction limit belong in general to this class of
junctions. Note that, if the length and width of the junction are $\ll\xi$, it
is irrelevant whether it is made from a normal metal or from a superconductor.
In the literature, such a junction is referred to as an ScS junction, without
specifying whether the constriction (c) is N or S.
\item Alternatively, the step-function model holds if the resistivity of the
junction region is much bigger than the resistivity of the bulk superconductor.
This condition is formulated more precisely by Kupriyanov et al. \cite{Kup82},
who have also studied deviations from the ideal step-function model due to the
proximity effect and due to current-induced suppression of the pair potential.
A superconductor --- semiconductor --- superconductor junction is typically in
this second category.
\end{enumerate}

In equilibrium, a phase difference $\phi$ in the pair potential induces a
stationary current $I$ through the junction (DC Josephson effect). The
current--phase relationship $I(\phi)$ is $2\pi$-periodic and antisymmetric
[$I(\phi +2\pi)=I(\phi)$, $I(-\phi)=-I(\phi)$]. The maximum $I_{\rm
c}\equiv{\rm max}\, I(\phi)$ is known as the critical current. There is a
thermodynamic relation
\begin{eqnarray}
I=\frac{2e}{\hbar}\,\frac{dF}{d\phi}\label{dFdphi}
\end{eqnarray}
between the equilibrium current and the derivative of the free energy $F$
with respect to the phase difference. To apply this relation we need
to know how to obtain $F$ from the BdG equation. The required formula was
derived by Bardeen
et al.\ \cite{Bar69} from the Greens function expression for $F$. An
alternative derivation, directly from the BdG equation, has been given in Ref.\
{\cite{SFe91}. The result is
\begin{eqnarray}
F&=&-2k_{\rm B}T\sum_{\varepsilon>0}\ln\left[
2\cosh( \varepsilon/2k_{\rm B}T)\right] +\int\! d{\bf r}\,|\Delta|^{2}/|g|
+{\rm Tr}\,{\cal H}_{0} .\label{F7}
\end{eqnarray}
The first term in Eq.\ (\ref{F7}) (the sum over $\varepsilon$) can be
formally interpreted as the free energy of non-interacting electrons,
all of one single spin, in a ``semiconductor'' with Fermi level halfway
between the ``conduction band'' (positive $\varepsilon$) and the ``valence
band'' (negative $\varepsilon$). This semiconductor
model of a superconductor appeals to intuition, but does not give the
free energy correctly. The second term in
Eq.\ (\ref{F7}) corrects for a double-counting of the interaction
energy in the semiconductor model. The third term ${\rm Tr}\,{\cal
H}_{0}$ (i.e.\ the sum of the single-electron eigenenergies) cancels a
divergence at large $\varepsilon$ of the series in the
first term.

{}From Eqs.\ (\ref{dFdphi}) and (\ref{F7}) one obtains \cite{SFe91}
\begin{eqnarray}
I&=&-\frac{2e}{\hbar}\sum_{p}\tanh(\varepsilon_{p}/2k_{\rm B}T)
\frac{d\varepsilon_{p}}{d\phi}\nonumber\\
&&\!\!\! -\frac{2e}{\hbar}2k_{\rm B}T
\int_{\Delta_{0}}^{\infty}\!d{\varepsilon}\,\ln\left[ 2\cosh(
\varepsilon/2k_{\rm B}T)\right]
\frac{d\rho}{d\phi}+\frac{2e}{\hbar}\,\frac{d}{d\phi} \int\!d{\bf
r}\,|\Delta|^{2}/|g| ,\label{dFdphi2}
\end{eqnarray}
where we have rewritten $\sum_{\varepsilon>0}$ as a sum over the discrete
positive eigenvalues $\varepsilon_{p}$ ($p=1,2,\ldots$), and an integration
over
the continuous spectrum with density of states $\rho(\varepsilon)$. The
term ${\rm Tr}\,{\cal H}_{0}$ in Eq.\ (\ref{F7}) does not depend on
$\phi$, and therefore does not contribute to $I$. The spatial integral of
$|\Delta|^{2}/|g|$ does contribute in general. If the step-function model for
$\Delta({\bf r})$ holds, however, $|\Delta|$ is independent of $\phi$ so that
this contribution can be disregarded. A calculation of the Josephson current
from Eq.\ (\ref{dFdphi2}) then requires only knowledge of the eigenvalues, not
of the eigenfunctions.

\section{Excitation spectrum from scattering matrix}

Following Ref.\ \cite{Bee91a}, we now show how to relate the excitation
spectrum of Bogoliubov quasiparticles to the scattering matrix of normal
electrons. The supercurrent then follows from Eq.\ (\ref{dFdphi2}).
The model considered is illustrated in Fig.\ 1. It consists of a
disordered normal region between two
superconducting regions ${\rm S}_{1}$ and ${\rm S}_{2}$.
The disordered region may or may not contain a geometrical
constriction. To obtain a well-defined scattering problem
we insert ideal (impurity-free) normal leads ${\rm N}_{1}$
and ${\rm N}_{2}$ to the left and right of the disordered
region. The leads should be long compared to the Fermi wavelength $\lambda_{\rm
F}$, but short compared to the coherence length $\xi_{0}$. The SN interfaces
are located at $x=\pm L/2$.
We assume that the only scattering in the superconductors consists of Andreev
reflection at the SN interfaces, i.e.\ we consider the case
that the disorder is contained entirely within the normal region.
The spatial separation of Andreev and normal scattering
is the key simplification which allows us to relate the supercurrent directly
to the normal-state scattering matrix.

\begin{figure}[tb]
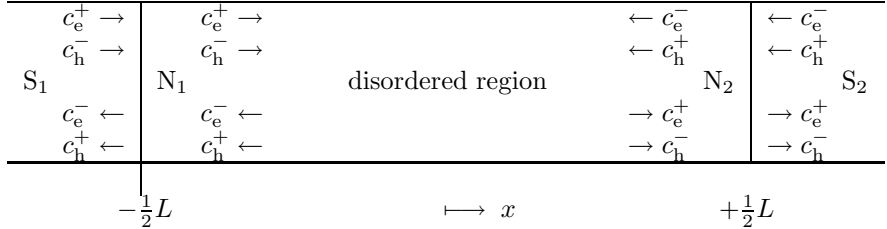

\vspace*{1cm}
\centering
\begin{tabular}{r|rp{4cm}l|l}
\hline
$c_{\rm e}^{+}\rightarrow$&$c_{\rm e}^{+}\rightarrow$&&$\leftarrow c_{\rm
e}^{-}$&$\leftarrow c_{\rm e}^{-}$\\
$c_{\rm h}^{-}\rightarrow$&$c_{\rm h}^{-}\rightarrow$&&$\leftarrow c_{\rm
h}^{+}$&$\leftarrow c_{\rm h}^{+}$\\
${\rm S}_{1}$\hspace*{1cm}&${\rm N}_{1}$\hspace*{1cm}&\hspace*{0.7cm}disordered
region&\hspace*{1cm}${\rm N}_{2}$&\hspace*{1cm}${\rm S}_{2}$\\
$c_{\rm e}^{-}\leftarrow$&$c_{\rm e}^{-}\leftarrow$&&$\rightarrow c_{\rm
e}^{+}$&$\rightarrow c_{\rm e}^{+}$\\
$c_{\rm h}^{+}\leftarrow$&$c_{\rm h}^{+}\leftarrow$&&$\rightarrow c_{\rm
h}^{-}$&$\rightarrow c_{\rm h}^{-}$\\
\hline\\
\multicolumn{5}{c}{$-\case{1}{2}L$\hspace*{3.6cm}$\longmapsto\;
x$\hspace*{2.7cm}$+\case{1}{2}L$}
\end{tabular}
\vspace*{1cm}
\caption{Superconductor --- normal-metal --- superconductor
Josephson junction containing a disordered normal region. Scattering states in
the normal and superconducting leads are indicated schematically. Normal
scattering (disordered region) and Andreev reflection (SN interface) are
spatially separated.}
\end{figure}

We first construct a basis for the scattering matrix ($s$-matrix).
In the normal lead ${\rm N}_{1}$ the eigenfunctions of the BdG equation
(\ref{BdG1}) can be written in the form
\begin{eqnarray}
&&\Psi_{n,{\rm e}}^{\pm}({\rm N}_{1})=
{\renewcommand{\arraystretch}{0.6}
\left(\begin{array}{c}1\\ 0\end{array}\right)}
(k_{n}^{\rm e})^{-1/2}\,\Phi_{n}(y,z)
\exp[\pm{\rm i}k_{n}^{\rm e}(x+\case{1}{2}L)],\nonumber\\
&&\Psi_{n,{\rm h}}^{\pm}({\rm N}_{1})=
{\renewcommand{\arraystretch}{0.6}
\left(\begin{array}{c}0\\ 1\end{array}\right)}
(k_{n}^{\rm h})^{-1/2}\,\Phi_{n}(y,z)
\exp[\pm{\rm i}k_{n}^{\rm h}(x+\case{1}{2}L)],\label{PsiN}
\end{eqnarray}
where the wavenumbers $k_{n}^{\rm e}$ and $k_{n}^{\rm h}$ are given by
\begin{eqnarray}
k_{n}^{\rm e,h}\equiv (2m/\hbar^{2})^{1/2}(E_{\rm F}
-E_{n}+\sigma^{\rm e,h}\varepsilon)^{1/2}, \label{keh}
\end{eqnarray}
and we have defined $\sigma^{\rm e}\equiv 1$,
$\sigma^{\rm h}\equiv -1$. The labels e and h indicate
the electron or hole character of the wavefunction.
The index $n$ labels
the modes, $\Phi_{n}(y,z)$ is the transverse wavefunction
of the $n$-th mode, and $E_{n}$ its threshold energy:
\begin{eqnarray}
[(p_{y}^{2}+p_{z}^{2})/2m+V(y,z)]\Phi_{n}(y,z)=E_{n}\Phi_{n}(y,z).\label{Phin}
\end{eqnarray}
The eigenfunction $\Phi_{n}$ is normalized to unity, $\int\! dy\int\! dz
\,|\Phi_{n}|^{2}=1$.
The eigenfunctions in lead ${\rm N}_{2}$ are chosen
similarly, but with $x+\case{1}{2}L$ replaced by $x-\case{1}{2}L$.

In the superconducting lead ${\rm S}_{1}$,
where $\Delta=\Delta_{0}\exp({\rm i}\phi/2)$,
the eigenfunctions are
\begin{eqnarray}
\Psi_{n,{\rm e}}^{\pm}({\rm S}_{1})=
{\renewcommand{\arraystretch}{0.6}
\left(\begin{array}{c}
{\rm e}^{{\rm i}\eta^{\rm\scriptstyle e}/2}\\
{\rm e}^{-{\rm i}\eta^{\rm\scriptstyle e}/2}\end{array}\right)}
(2q_{n}^{\rm e})^{-1/2}
(\varepsilon^{2}/\Delta_{0}^{2}-1)^{-1/4}
\Phi_{n}\exp[\pm{\rm i}q_{n}^{\rm e}(x+\case{1}{2}L)],\nonumber\\
\Psi_{n,{\rm h}}^{\pm}({\rm S}_{1})=
{\renewcommand{\arraystretch}{0.6}
\left(\begin{array}{c}
{\rm e}^{{\rm i}\eta^{\rm\scriptstyle h}/2}\\
{\rm e}^{-{\rm i}\eta^{\rm\scriptstyle h}/2}\end{array}\right)}
(2q_{n}^{\rm h})^{-1/2}
(\varepsilon^{2}/\Delta_{0}^{2}-1)^{-1/4}
\Phi_{n}\exp[\pm{\rm i}q_{n}^{\rm h}(x+\case{1}{2}L)].
\label{PsiS}
\end{eqnarray}
We have defined
\begin{eqnarray}
q_{n}^{\rm e,h}&\equiv &(2m/\hbar^{2})^{1/2}
[E_{\rm F}-E_{n}+\sigma^{\rm e,h}
(\varepsilon^{2}-\Delta_{0}^{2}) ^{1/2}]^{1/2},\label{qneh}\\
\eta^{\rm e,h}&\equiv &\case{1}{2}\phi+\sigma^{\rm e,h}
\arccos(\varepsilon/\Delta_{0}).\label{etaeh}
\end{eqnarray}
The square roots are to be taken such that ${\rm Re}\,
q^{\rm e,h}\geq 0$, ${\rm Im}\,q^{\rm e}\geq 0$,
${\rm Im}\,q^{\rm h}\leq 0$. The function $\arccos t\in(0,\pi/2)$
for $0<t<1$, while $\arccos t\equiv -{\rm i}
\ln[t+(t^{2}-1)^{1/2}]$
for $t>1$. The eigenfunctions in lead ${\rm S}_{2}$, where
$\Delta=\Delta_{0}\exp(-{\rm i}\phi/2)$,
are obtained by replacing $\phi$ by $-\phi$ and $L$ by $-L$.

The wavefunctions (\ref{PsiN}) and (\ref{PsiS}) have
been normalized to carry the same amount of quasiparticle
current, because we want to use them as the basis for a unitary
$s$-matrix. Unitarity of the $s$-matrix is a consequence of the fact that the
BdG equation conserves the quasiparticle current density
\begin{eqnarray}
{\bf j}_{\rm qp}({\bf r})=\frac{1}{m}{\rm Re}\,
{\renewcommand{\arraystretch}{0.8}
\left(\begin{array}{c}
{\rm u}^{\ast}\\
{\rm v}^{\ast}
\end{array}\right)
\left(\begin{array}{cc}
{\bf p}&\emptyset\\
\emptyset &-{\bf p}
\end{array}\right)
\left(\begin{array}{c}
{\rm u}\\
{\rm v}
\end{array}\right)
}.\label{jqp}
\end{eqnarray}
The quasiparticle current $I_{\rm qp}$ through a lead is given by
\begin{eqnarray}
I_{\rm qp}=\int\!\! dy\int\!\! dz\, ({\bf j}_{\rm qp})_{x}=\frac{\hbar}{m}{\rm
Re}\int\!\! dy\int\!\! dz\left( {\rm u}^{\ast}\frac{\partial}{{\rm i}\partial
x}{\rm u}-{\rm v}^{\ast}\frac{\partial}{{\rm i}\partial x}{\rm v}\right)
.\label{Iqp}
\end{eqnarray}
One easily verifies that, provided the wavenumbers $k_{n}^{\rm e,h}$ and
$q_{n}^{\rm e,h}$ are real (i.e.\ for propagating modes), each of the
wavefunctions (\ref{PsiN},\ref{PsiS}) carries the same amount of quasiparticle
current $|I_{\rm qp}|$. (The direction of the current is opposite for
$\Psi_{\rm e}^{+},\Psi_{\rm h}^{-}$ and $\Psi_{\rm e}^{-},\Psi_{\rm h}^{+}$.)
We note that the wavefunctions (\ref{PsiN}) and (\ref{PsiS}) carry {\em
different\/} amounts of {\em charge\/} current (i.e.\ electrical current),
which is given by
\begin{eqnarray}
I_{\rm charge}
=-e\frac{\hbar}{m}{\rm Re}\int\!\! dy\int\!\! dz\left( {\rm
u}^{\ast}\frac{\partial}{{\rm i}\partial x}{\rm u}+{\rm
v}^{\ast}\frac{\partial}{{\rm i}\partial x}{\rm v}\right) .\label{Icharge}
\end{eqnarray}
This does not contradict the unitarity of the $s$-matrix constructed from these
wavefunctions, because the BdG equation does not conserve the charge of the
quasiparticles. (Andreev reflection \cite{And64}, i.e.\ the reflection of an
electron into a hole at an SN interface, is an example of a
non-charge-conserving scattering process.)

A wave incident on the disordered normal
region is described in the basis (\ref{PsiN})
by a vector of coefficients
\begin{eqnarray}
c_{\rm N}^{\rm in}\equiv\left(
c_{\rm e}^{+}(N_{1}), c_{\rm e}^{-}(N_{2}),
c_{\rm h}^{-}(N_{1}), c_{\rm h}^{+}(N_{2})\right),\label{cNin}
\end{eqnarray}
as shown schematically in Fig.\ 1.
(The mode-index $n$ has been suppressed
for simplicity of notation.)
The reflected and transmitted wave has vector of coefficients
\begin{eqnarray}
c_{\rm N}^{\rm out}\equiv\left(
c_{\rm e}^{-}(N_{1}), c_{\rm e}^{+}(N_{2}),
c_{\rm h}^{+}(N_{1}), c_{\rm h}^{-}(N_{2})\right).\label{cNout}
\end{eqnarray}
The $s$-matrix $s_{\rm N}$ of the normal region
relates these two vectors, $c_{\rm N}^{\rm out}=
s_{\rm N}c_{\rm N}^{\rm in}$. Because the normal
region does not couple electrons and holes, this
matrix has the block-diagonal form
\begin{eqnarray}
s_{\rm N}(\varepsilon)=
{\renewcommand{\arraystretch}{0.6}
\left(\begin{array}{cc}
s_{0}(\varepsilon)&\emptyset\\
\!\emptyset&s_{0}(-\varepsilon)^{\ast}
\end{array}\right)},\,
s_{\rm 0}\equiv{\renewcommand{\arraystretch}{0.6}
\left(\begin{array}{cc}
r_{11}&t_{12}\\t_{21}&r_{22}
\end{array}\right)}.
\label{sN}
\end{eqnarray}
Here $s_{0}$ is the unitary and symmetric $s$-matrix associated
with the single-electron Hamiltonian ${\cal H}_{0}$.
The reflection and transmission matrices $r(\varepsilon)$ and $t(\varepsilon)$
are $N\times N$ matrices, $N(\varepsilon)$ being the number of
propagating modes at energy $\varepsilon$. (We assume
for simplicity that the number of modes in
leads $N_{1}$ and $N_{2}$ is the same.) The dimension of $s_{\rm
N}(\varepsilon)$ is $2N(\varepsilon)+2N(-\varepsilon)$.

We will make use of two more $s$-matrices.
For energies $0<\varepsilon<\Delta_{0}$ there are
no propagating modes in the superconducting leads
${\rm S}_{1}$ and ${\rm S}_{2}$. We can then define
an $s$-matrix $s_{\rm A}$ for Andreev
reflection at the SN interfaces by $c_{\rm N}^{\rm in}
=s_{\rm A}c_{\rm N}^{\rm out}$. The elements of
$s_{\rm A}$ can be obtained by
matching the wavefunctions (\ref{PsiN}) at $|x|=L/2$ to the
decaying wavefunctions (\ref{PsiS}). Since
$\Delta_{0}\ll E_{\rm F}$ one may ignore
normal reflections at the SN interface and neglect the difference between
$N(\varepsilon)$ and $N(-\varepsilon)$. This is known as the Andreev
approximation \cite{And64}. The result is
\begin{eqnarray}
s_{\rm A}=\alpha
{\renewcommand{\arraystretch}{0.6}
\left(\begin{array}{cc}
\emptyset&r_{\rm A}\\
r_{\rm A}^{\ast}&\emptyset
\end{array}\right)},\,
r_{\rm A}\equiv
{\renewcommand{\arraystretch}{0.6}
\left(\begin{array}{cc}
{\rm e}^{{\rm i}\phi/2}\one&\emptyset\\
\emptyset&{\rm e}^{-{\rm i}\phi/2}\one
\end{array}\right)},
\label{sA}
\end{eqnarray}
where $\alpha\equiv\exp[-{\rm i}\arccos(\varepsilon/\Delta_{0})]$.
The matrices $\one$ and $\emptyset$
are the unit and null matrices, respectively.
Andreev reflection transforms an electron mode into a hole mode, without change
of mode index. The transformation is accompanied by a phase shift, which
consists of two parts: 1. A phase shift $-\arccos(\varepsilon/\Delta_{0})$ due
to the penetration of the wavefunction into the superconductor; 2. A phase
shift equal to plus or minus the phase of the pair potential in the
superconductor ({\em plus\/} for reflection from hole to electron, {\em
minus\/} for the reverse process).

For $\varepsilon>\Delta_{\rm 0}$ we can define the
$s$-matrix $s_{\rm SNS}$ of the whole junction,
by $c_{\rm S}^{\rm out}=s_{\rm SNS}c_{\rm S}^{\rm in}$.
The vectors
\begin{eqnarray}
c_{\rm S}^{\rm in}&\equiv&\left(
c_{\rm e}^{+}(S_{1}), c_{\rm e}^{-}(S_{2}),
c_{\rm h}^{-}(S_{1}), c_{\rm h}^{+}(S_{2})\right),\label{cSin}\\
c_{\rm S}^{\rm out}&\equiv&\left(
c_{\rm e}^{-}(S_{1}), c_{\rm e}^{+}(S_{2}),
c_{\rm h}^{+}(S_{1}), c_{\rm h}^{-}(S_{2})\right).\label{cSout}
\end{eqnarray}
are the coefficients in the expansion of the
incoming and outgoing wave in leads ${\rm S}_{1}$
and ${\rm S}_{2}$ in terms of the wavefunctions (\ref{PsiS}) (cf.\ Fig.\ 1).
By matching the wavefunctions (\ref{PsiN}) and (\ref{PsiS}) at
$|x|=L/2$, we arrive after some algebra (using again $\Delta_{0}\ll E_{\rm F}$)
at the matrix-product expression
\begin{eqnarray}
&&s_{\rm SNS}=U^{-1}\,(\one-M)^{-1}\,
(\one-M^{\dagger})
s_{\rm N}U,\label{sSNS}\\
&&U\equiv
{\renewcommand{\arraystretch}{0.6}
\left(\begin{array}{cc}
r_{\rm A}&\emptyset\\
\emptyset&r_{\rm A}^{\ast}
\end{array}\right) } ^{\!\frac{1}{2}} \! ,\;
M\equiv\alpha s_{\rm N}
{\renewcommand{\arraystretch}{0.6}
\left(\begin{array}{cc}
\emptyset&r_{\rm A}\\
r_{\rm A}^{\ast}&\emptyset
\end{array}\right)}.\nonumber
\end{eqnarray}

One can verify that the three $s$-matrices defined above
($s_{\rm N}$; $s_{\rm A}$ for $0<\varepsilon <\Delta_{0}$;
$s_{\rm SNS}$ for $\varepsilon >\Delta_{0}$) are
{\em unitary\/} ($s^{\dagger}s=ss^{\dagger}=\one$) and satisfy the {\em
symmetry relation\/} $s(\varepsilon,\phi)_{ij}=
s(\varepsilon,-\phi)_{ji}$, as required by quasiparticle-current conservation
and by time-reversal invariance, respectively.

We are now ready to relate the excitation spectrum
of the Josephson junction to the $s$-matrix of the normal region.
First the discrete spectrum. The condition $c_{\rm in}=
s_{\rm A}s_{\rm N}c_{\rm in}$ for a bound state implies
${\rm Det}\,(\one-s_{\rm A}s_{\rm N})=0$. Using
Eqs.\ (\ref{sN}), (\ref{sA}), and the {\em folding-identity}
\begin{eqnarray}
{\rm Det}\,
{\renewcommand{\arraystretch}{0.4}
\left(\begin{array}{cc}
a&b\\c&d
\end{array}\right)}={\rm Det}\,(ad-aca^{-1}b)
\label{identity}
\end{eqnarray}
(which holds for arbitrary square matrices $a,b,c,d$ of equal dimension, with
${\rm Det}\, a\neq 0$), we find the equation
\begin{eqnarray}
{\rm Det}\,\left[\one-\alpha(\varepsilon_{p})^{2}
r_{\rm A}^{\ast}s_{0}(\varepsilon_{p})r_{\rm A}
s_{0}(-\varepsilon_{p})^{\ast}\right] =0,\label{discrete}
\end{eqnarray}
which determines the discrete spectrum.
The density of states of the
continuous spectrum is related to
$s_{\rm SNS}$ by the general relation \cite{Akk91}
\begin{eqnarray}
\rho=\frac{1}{2\pi{\rm i}}\,\frac{\partial}{\partial\varepsilon}
\ln{\rm Det}\, s_{\rm SNS}+{\rm constant},\label{rho}
\end{eqnarray}
where ``constant'' indicates a $\phi$-independent term. From Eqs.\ (\ref{sSNS})
and (\ref{identity}) we find
\begin{eqnarray}
\frac{\partial\rho}{\partial\phi}
=-\frac{1}{\pi}\,
\frac{\partial^{2}}{\partial\phi\partial\varepsilon}
{\rm Im}\ln{\rm Det}
\left[\one-\alpha(\varepsilon)^{2}
r_{\rm A}^{\ast}s_{0}(\varepsilon)r_{\rm A}
s_{0}(-\varepsilon)^{\ast}\right] ,
\label{cont}
\end{eqnarray}
which determines the $\phi$-dependence of the continuous spectrum.

\section{Short-Junction Limit}

In the short-junction limit $L\ll\xi$, the determinantal equations
(\ref{discrete}) and (\ref{cont})
can be simplified further. As mentioned in the Introduction, the condition
$L\ll\xi$ is
equivalent to $\Delta_{0}\ll E_{\rm c}$, where the
correlation energy
$E_{\rm c}\equiv\hbar/\tau_{\rm dwell}$ is defined in terms of the
dwell time $\tau_{\rm dwell}$
in the junction. The elements of
$s_{0}(\varepsilon)$
change significantly if $\varepsilon$ is changed by
at least $E_{\rm c}$ \cite{Sto86,Imr86a}. We are concerned with
$\varepsilon$ of order $\Delta_{0}$ or smaller (since
$\rho(\varepsilon,\phi)$ becomes independent
of $\phi$ for $\varepsilon
\gg\Delta_{0}$). For $\Delta_{0}\ll E_{\rm c}$ we may thus
approximate $s_{0}(\varepsilon)\approx s_{0}(-\varepsilon)\approx
s_{0}(0)\equiv s_{0}$. Eq.\ (\ref{discrete}) may now be simplified by
multiplying both sides by ${\rm Det}\,s_{0}$ and using $s_{0}^{\ast}s_{0}=\one$
(unitarity plus symmetry of $s_{0}$), as well as the folding identity
(\ref{identity}). The result can be written in the form
\begin{eqnarray}
{\rm Det}\left[ (1-\varepsilon_{p}^{2}/\Delta_{0}^{2})
\one-t_{12}t_{12}^{\dagger}
\sin^{2}(\phi/2)\right] =0.
\label{discrete2}
\end{eqnarray}
For $\varepsilon >\Delta_{0}$ one can see that ${\rm Det}
[\one-\alpha(\varepsilon)^{2}
r_{\rm A}^{\ast}s_{0}r_{\rm A}s_{0}^{\ast}]$ is a real number. (Use that
$\alpha$ is real for $\varepsilon >\Delta_{0}$ and that the determinant ${\rm
Det}\,(\one-aa^{\ast})$ is real for arbitrary matrix $a$.)
Eq.\ (\ref{cont}) then reduces to
${\partial\rho}/{\partial\phi}=0$, from which we conclude that {\em the
continuous spectrum does not contribute to
$I(\phi)$ in the short-junction limit}.

Eq.\ (\ref{discrete2}) can be solved for $\varepsilon_{p}$
in terms of the eigenvalues $T_{p}$ ($p=1,2,\ldots N$)
of the hermitian $N\times N$
matrix $t_{12}t_{12}^{\dagger}$,
\begin{eqnarray}
\varepsilon_{p}=\Delta_{0}\left[ 1-T_{p}\sin^{2}
(\phi/2)\right] ^{1/2}.\label{discrete3}
\end{eqnarray}
Since $t_{12}t_{12}^{\dagger}=r_{11}t_{21}^{\dagger}t_{21}r_{11}^{-1}$ (as
follows from unitarity of $s_{0}$), the matrices $t_{12}t_{12}^{\dagger}$ and
$t_{21}t_{21}^{\dagger}$ have the same set of eigenvalues. We can therefore
omit the indices of $tt^{\dagger}$. Unitarity of $s_{0}$ also implies that
$0\leq T_{p}\leq 1$ for all $p$.

Substitution of Eq.\ (\ref{discrete3}) into Eq.\ (\ref{dFdphi2}) yields the
Josephson current
\begin{eqnarray}
I(\phi )=\frac{e\Delta_{0}}{2\hbar}\sum_{p=1}^{N}
\frac{T_{p}\sin\phi}{[1-T_{p}\sin^{2}(\phi/2)]^{1/2}}
\tanh\left( \frac{\Delta_{0}}{2k_{\rm
B}T}[1-T_{p}\sin^{2}(\phi/2)]^{1/2}\right) .\label{Iphi2}
\end{eqnarray}
Eq.\ (\ref{Iphi2}) holds for an arbitrary
transmission matrix $tt^{\dagger}$, i.e.\ for arbitrary disorder potential.
It is the {\em multi-channel\/} generalization of a formula first obtained by
Haberkorn et al.\ \cite{Hab78} (and subsequently rederived by several authors
\cite{Zai84}--\cite{Fur90})
for the {\em single-channel\/} case (appropriate for a geometry such as a
planar tunnel barrier, where the different scattering channels are uncoupled).
A formula of similar generality for the conductance is the multi-channel
Landauer formula \cite{Imr86a,Fis81}
\begin{eqnarray}
G=\frac{2e^{2}}{h}{\rm Tr}\,tt^{\dagger}\equiv\frac{2e^{2}}{h}
\sum_{p=1}^{N}T_{p}.\label{Landauer}
\end{eqnarray}
In contrast to the Landauer formula, Eq.\ (\ref{Iphi2}) is a {\em non-linear\/}
function of the transmission eigenvalues $T_{p}$. It follows that knowledge of
the conductance (i.e.\ of the sum of the eigenvalues) is {\em not sufficient\/}
to determine the supercurrent.

\section{Universal Josephson Effects}

\subsection{Quantum Point Contact}

Consider the case that the weak link consists of a
ballistic constriction ($l\gg L$) with
a conductance quantized at $G=2N_{0}e^{2}/h$ (a {\em quantum point contact\/}
\cite{Eer91}). The integer $N_{0}$ is the number of occupied one-dimensional
subbands (per spin direction) in the constriction, or alternatively the number
of transverse modes at the Fermi level which can propagate through the
constriction. Note that $N_{0}\ll N$. A quantum point contact is characterized
by a special set of transmission eigenvalues, which are equal to either 0 or 1:
\begin{eqnarray}
T_{p}=\left\{\begin{array}{ll}
1 &\;{\rm if}\; 1\leq p\leq N_{0},\\
0 &\;{\rm if}\; N_{0}<p\leq N,
\end{array}\right.\label{TQPC}
\end{eqnarray}
where the eigenvalues have been ordered from large to small. We emphasize that
Eq.\ (\ref{TQPC}) is valid whether the transport through the constriction is
adiabatic or not. In the special case of adiabatic transport, the transmission
eigenvalue $T_{p}$ is equal to the transmission probability ${\cal T}_{p}$ of
the $p$-th subband. In the absence of adiabaticity there is no simple relation
between $T_{p}$ and ${\cal T}_{p}$.

The discrete spectrum (\ref{discrete3}) in the short-junction limit contains an
$N_{0}$-fold degenerate state at energy $\varepsilon=\Delta_{0}|\cos(\phi/2)|$.
Eq.\ (\ref{Iphi2}) for the supercurrent becomes
\begin{eqnarray}
I(\phi)=N_{0}\frac{e\Delta_{0}}{\hbar}
\sin(\phi/2)\tanh\left(\frac{\Delta_{0}} {2k_{\rm
B}T}\cos(\phi/2)\right),\; |\phi|<\pi .\label{IQPC}
\end{eqnarray}
At $T=0$ the current--phase relationship is given by
$I(\phi)=N_{0}(e\Delta_{0}/\hbar)\sin(\phi/2)$ for $|\phi |<\pi$, and continued
periodically for $|\phi |>\pi$. At $\phi =\pi$ the function $I(\phi )$ has a
discontinuity which is smeared at finite temperatures. The ratio $I(\phi
)/(\pi\Delta_{0}G/e)$ at $T=0$ is plotted in Fig.\ 2 (solid curve). The
critical current $I_{\rm c}=N_{0}e\Delta_{0}/\hbar$ is {\em discretized\/} in
units of $e\Delta_{0}/\hbar$. In the classical limit $N_{0}\rightarrow\infty$
we recover the results of Kulik and Omel'yanchuk \cite{Kul77} for a {\em
classical \/} ballistic point contact.

\begin{figure}[tb]
\centerline{
\psfig{figure=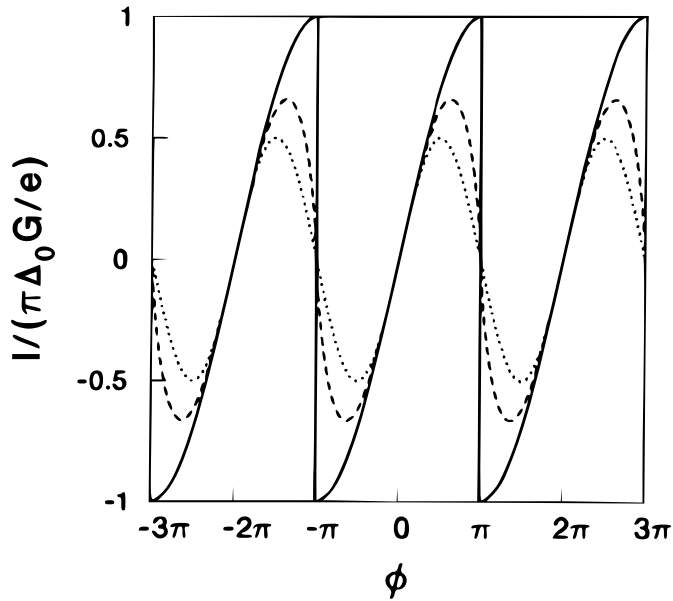,width= .8\linewidth}}
\caption{Current--phase relationship at $T=0$ of a ballistic
point contact [solid curve,  Eq.\ (\protect\ref{IQPC})], a disordered point
contact (ensemble averaged) [dashed curve, Eq.\ (\protect\ref{Iav2})], and
a tunnel junction [dotted curve, Eq.\ (\protect\ref{Iphi3})]. The current is
normalized by the normal-state conductance of the junction.}
\end{figure}

Eq.\ (\ref{IQPC}) was derived
in Ref.\ \cite{Bee91} under the assumption of adiabatic transport. The present
derivation (taken from Ref.\ \cite{Bee91a}) does not assume adiabaticity. As
discussed in Ref.\ \cite{Bee91}, Eq.\ (\ref{IQPC}) breaks down if the Fermi
level lies within $\Delta_{0}$ of the threshold energy $E_{n}$ of a subband. In
the present context this breakdown can be understood from the fact that the
dwell time $\tau_{\rm dwell}=(L/v_{\rm F})(1-E_{n}/E_{\rm F})^{-1/2}$ becomes
infinitely long as $E_{\rm F}$ approaches $E_{n}$, so that the
``short-junction'' condition $\Delta_{0}\ll E_{\rm c}\equiv\hbar/\tau_{\rm
dwell}$ is violated.

\subsection{Quantum Dot}

Consider a small confined region (of dimensions comparable to the Fermi
wavelength), which is weakly coupled by tunnel barriers to two electron
reservoirs. We assume that transport through this {\em quantum dot\/} occurs
via resonant tunneling through a single
bound state. Let $\varepsilon_{\rm res}$ be the energy of the resonant
level, relative to the Fermi energy $E_{\rm F}$ in the reservoirs, and
let $\Gamma_{1}/\hbar$ and $\Gamma_{2}/\hbar$ be the tunnel rates
through the left and right barriers. We denote
$\Gamma\equiv\Gamma_{1}+\Gamma_{2}$. If $\Gamma\ll\Delta E$ (with $\Delta E$
the level spacing in the quantum dot) and $T\ll\Gamma/k_{\rm B}$, the
conductance $G$ in the case of non-interacting electrons has the form
\begin{eqnarray}
G=\frac{2e^2}{h}\,\frac{\Gamma_{1}\Gamma_{2}}{\varepsilon_{\rm
res}^{2}+{\case{1}{4}}\Gamma^{2}}\equiv\frac{2e^{2}}{h}T_{\rm
BW},\label{GBW}
\end{eqnarray}
where $T_{\rm BW}$ is the Breit-Wigner
transmission probability at the Fermi level. The normal-state scattering matrix
 $s_{0}(\varepsilon)$ which yields this conductance has matrix elements
\cite{LL77}
\begin{eqnarray}
\left( s_{0}(\varepsilon)\right) _{nm}=\left( \delta_{nm}-\frac{{\rm
i}\sqrt{\Gamma_{1n}\Gamma_{2m}}}{\varepsilon-\varepsilon_{\rm res}+{\rm
i}\Gamma/2}\right) {\rm e}^{{\rm i}(\delta_{1n}+\delta_{2m})} ,\label{sBW}
\end{eqnarray}
where $\sum_{n=1}^{N}\Gamma_{1n}\equiv\Gamma_{1}$,
$\sum_{n=1}^{N}\Gamma_{2n}\equiv\Gamma_{2}$. The phases $\delta_{1n}$,
$\delta_{2m}$, as well as the basis with respect to which the matrix elements
(\ref{sBW}) are calculated, need not be further specified.
B\"{u}ttiker \cite{But88} has shown how the conductance (\ref{GBW}) follows,
via the Landauer formula (\ref{Landauer}), from the Breit-Wigner scattering
matrix (\ref{sBW}).

\begin{figure}[tb]
\centerline{
\psfig{figure=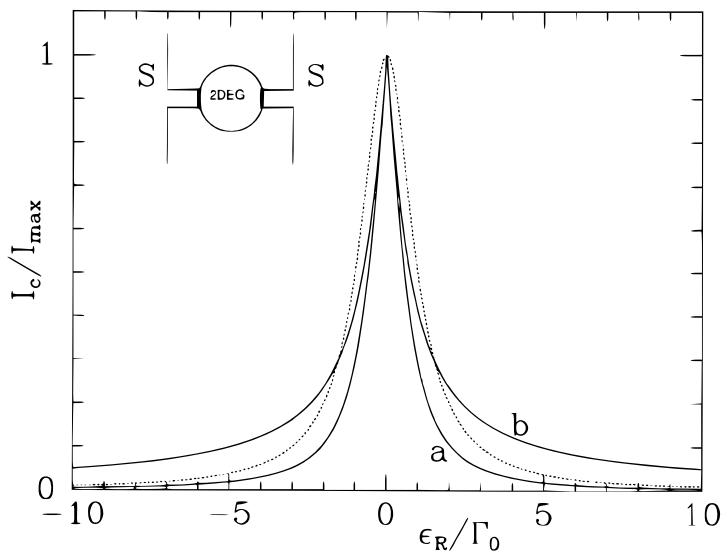,width= .8\linewidth}}
\caption{Normalized critical current versus energy of the resonant
level, at zero temperature and for equal tunnel barriers
($\Gamma_{1}=\Gamma_{2}\equiv\Gamma_{0}$). The two solid curves are the
results (\protect\ref{Icwide}) and (\protect\ref{Icnarrow}) for the two
regimes $\Gamma_{0}\gg\Delta_{0}$ (curve a) and
$\Gamma_{0},\varepsilon_{\rm res}\ll\Delta_{0}$ (curve b). The dotted curve
is the Breit-Wigner transmission probability (\protect\ref{GBW}). The
inset shows schematically the quantum-dot Josephson junction. Taken from Ref.\
\protect\cite{SQUID91}.}
\end{figure}

In Ref.\ \cite{SQUID91} Van Houten and the author have calculated the
supercurrent through the quantum dot from the Breit-Wigner formula (\ref{sBW})
in the special case of single-channel leads ($N=1$). Just as for the
conductance \cite{But88}--\cite{Xue88} one would expect that the results are
not changed for multi-channel leads ($N>1$). We will now demonstrate this
explicitly in the limit $\Gamma\gg\Delta_{0}$ (which for resonant tunneling
corresponds to the ``short-junction'' limit, cf.\ Sec.\ 1). The transmission
matrix product $t_{12}t_{12}^{\dagger}$ (evaluated at the Fermi level
$\varepsilon=0$) following from the scattering matrix (\ref{sBW}) is
\begin{eqnarray}
\left( t_{12}t_{12}^{\dagger}\right) _{nm}=T_{\rm
BW}\gamma_{n}\gamma_{m}^{\ast},\; \gamma_{n}\equiv{\rm e}^{{\rm
i}\delta_{n}}\sqrt{\Gamma_{1n}/\Gamma_{1}}.\label{ttBW}
\end{eqnarray}
Its eigenvalues are
\begin{eqnarray}
T_{p}=\left\{\begin{array}{ll}
T_{\rm BW} &\;{\rm if}\; p=1,\\
0 &\;{\rm if}\; 2\leq p\leq N.
\end{array}\right.\label{TBW}
\end{eqnarray}
The single non-zero eigenvalue has eigenvector $u_{n}=\gamma_{n}$, the other
eigenvectors span the plane orthogonal to the vector $\gamma$. Substitution
into Eq.\ (\ref{Iphi2}) yields the current--phase relationship for a wide
resonance ($\Gamma\gg\Delta_{0})$,
\begin{eqnarray}
I(\phi )=\frac{e\Delta_{0}}{2\hbar}
\frac{T_{\rm BW}\sin\phi}{[1-T_{\rm BW}\sin^{2}(\phi/2)]^{1/2}}
\tanh\left( \frac{\Delta_{0}}{2k_{\rm B}T}[1-T_{\rm
BW}\sin^{2}(\phi/2)]^{1/2}\right) .\label{IBW}
\end{eqnarray}
The critical current at zero temperature is
\begin{eqnarray}
I_{\rm c}=\frac{e\Delta_{0}}{\hbar}[1-(1-T_{\rm
BW})^{1/2}],\;{\rm if}\;\Gamma\gg\Delta_{0},\label{Icwide}
\end{eqnarray}
in agreement with Ref.\ \cite{SQUID91}. Since $T_{\rm BW}=1$ on resonance
($\varepsilon_{\rm R}=0$) in the case of equal tunnel rates
($\Gamma_{1}=\Gamma_{2}$), we obtain the result (\ref{effect2b}) discussed in
the Introduction.

For completeness, we also quote the formula for the current--phase relationship
in the opposite regime of a narrow resonance (derived in Ref.\ \cite{SQUID91}):
\begin{eqnarray}
I_{\rm
c}=\frac{e}{\hbar}(\varepsilon_{\rm
res}^{2}+{\case{1}{4}}\Gamma^{2})^{1/2}\,[1-(1-T_{\rm
BW})^{1/2}],\;{\rm if}\;\Gamma ,\varepsilon_{\rm
res}\ll\Delta_{0}.\label{Icnarrow}
\end{eqnarray}

As shown in Fig.\ 3, the lineshapes (\ref{Icwide})
and (\ref{Icnarrow}) of a resonance in the critical current (solid
curves) differ substantially from the lorentzian lineshape (\ref{GBW})
of a conductance resonance (dotted curve). For $\Gamma_{1}=\Gamma_{2}$,
$I_{\rm c}$ has a {\em cusp\/} at $\varepsilon_{\rm res}=0$ (which is
rounded at finite temperatures). On resonance,
the maximum critical current $I_{\rm c}^{\rm res}$ equals
$(2e\Delta_{0}/\hbar\Gamma)\,{\rm min}\,(\Gamma_{1},\Gamma_{2})$ and
$(e/\hbar)\,{\rm min}\,(\Gamma_{1},\Gamma_{2})$ for a wide and narrow
resonance, respectively. An analytical formula, Eq.\ (\ref{effect2a}), for the
crossover between these two regimes can be obtained for the case of equal
tunnel
rates \cite{SQUID91}. Off-resonance, $I_{\rm c}$
has the lorentzian decay $\propto 1/\varepsilon_{\rm res}^{2}$ in the case
$\Gamma\gg\Delta_{0}$ of a wide resonance, but a slower decay $\propto
1/\varepsilon_{\rm res}$ in the case $\Gamma ,\varepsilon_{\rm
res}\ll\Delta_{0}$. Near $\varepsilon_{\rm res}\simeq\Delta_{0}$ this linear
decay of the narrow resonance crosses over to a quadratic decay (not
shown in Fig.\ 3).

Since we have assumed non-interacting quasiparticles, the above
results apply to a quantum dot with a small charging energy $U$ for
double occupancy of the resonant state. Glazman and Matveev have
studied the influence of Coulomb repulsion on the resonant supercurrent
\cite{Gla89}. The influence is most pronounced in the case of a
narrow resonance, when the critical current is suppressed by a factor
$\Gamma/\Delta_{0}$ (for $U,\Delta_{0}\gg\Gamma$).  In the case of a
wide resonance, the Coulomb repulsion does not suppress the supercurrent, but
slightly broadens the resonance by a factor
$\ln(\Gamma/\Delta_{0})$ (for $U,\Gamma\gg\Delta_{0}$). The broadening
is a consequence of the Kondo effect, and occurs only for
$\varepsilon_{\rm res}<0$, so that the resonance peak becomes somewhat
asymmetric \cite{Gla89}.

\subsection{Disordered Point Contact}

We now turn to the regime of diffusive transport through a disordered point
contact. We first consider the average supercurrent (averaged over an ensemble
of impurity configurations) and then the fluctuations from the average. This
section corrects Ref.\ \cite{Bee91a}.

\subsubsection{Average Supercurrent}

The transmission eigenvalue $T_{p}$ is related to a channel-dependent
localization length $\zeta_{p}$ by
\begin{equation}
T_{p}=\cosh^{-2}(L/\zeta_{p}).\label{xip}
\end{equation}
The inverse localization length is uniformly distributed between $0$ and
$1/\zeta_{\rm min}\simeq 1/l$ for $l\ll L\ll Nl$ \cite{Sto91}. One can
therefore write
\begin{eqnarray}
\frac{\left\langle\sum_{p=1}^{N}f(T_{p})\right\rangle}
{\left\langle\sum_{p=1}^{N}T_{p}\right\rangle}&=&\frac{\int_{0}^{L/\zeta_{\rm
min}}\! dx \, f(\cosh^{-2}x)}{\int_{0}^{L/\zeta_{\rm min}}\! dx \,
\cosh^{-2}x}=\int_{0}^{\infty}\! dx\, f(\cosh^{-2}x),\label{avf}
\end{eqnarray}
where $\langle\ldots\rangle$ indicates the ensemble average and $f(T)$ is an
arbitrary function of the transmission eigenvalue such that $f(T)\rightarrow 0$
for $T\rightarrow 0$. In the second equality in Eq.\ (\ref{avf}) we have used
that $L/\zeta_{\rm min}\simeq L/l\gg 1$ to replace the upper integration limit
by $\infty$. Note that, in view of the Landauer formula (\ref{Landauer}), the
denominator $\left\langle\sum_{p=1}^{N}T_{p}\right\rangle$ is just $h/2e^{2}$
times the average conductance $\langle G\rangle$.

Combining Eqs.\ (\ref{avf})
and (\ref{Iphi2}) we find for the average supercurrent the expression
($0<\phi<\pi$)
\begin{eqnarray}
\langle I(\phi)\rangle &=&\frac{e\Delta_{0}}{2\hbar}\,\frac{h}{2e^{2}}\langle
G\rangle\int_{0}^{\infty}\! dx \,
\frac{(\cosh x)^{-2}\sin\phi}{[1-(\cosh
x)^{-2}\sin^{2}(\phi/2)]^{1/2}}\nonumber\\
&&\mbox{}\times\tanh\left( \frac{\Delta_{0}}{2k_{\rm B}T}[1-(\cosh
x)^{-2}\sin^{2}(\phi/2)]^{1/2}\right) \nonumber\\
&=& \frac{\pi\Delta_{0}}{e}\langle
G\rangle\cos(\phi/2)\int_{\cos(\phi/2)}^{1}\!
dz\,\frac{\tanh(\Delta_{0}z/2k_{\rm B}T)}{[z^{2}-\cos^{2}(\phi/2)]^{1/2}}
.\label{Iav1}
\end{eqnarray}
At $T=0$ this integral can be evaluated in closed form, with the result
\begin{eqnarray}
\langle I(\phi)\rangle =
\frac{\pi\Delta_{0}}{e}\langle G\rangle\cos(\phi/2)\,{\rm
arctanh}\left[\sin(\phi/2)\right] ,\label{Iav2}
\end{eqnarray}
plotted in Fig.\ 2 (dashed curve). Note that the derivative $dI/d\phi$ diverges
at $\phi =\pi$. The critical current is $I_{\rm c}^{(0)}\equiv{\rm
max}\,\langle I(\phi)\rangle=1.32\,(\pi\Delta_{0}
/2e)\langle G\rangle$, reached at $\phi\equiv\phi_{\rm c}^{(0)}=1.97$.

Eq.\ (\ref{Iav2}) for the average supercurrent in a disordered point contact
agrees with the result of Kulik and Omel'yanchuk \cite{Kul75}. These authors
pointed out that a disordered weak link and a tunnel junction with the {\em
same\/} conductance exhibit nevertheless a quite {\em different\/} Josephson
effect. For a tunnel junction with $T_{p}\ll 1$ for all $p$ one may linearize
Eq.\ (\ref{Iphi2}) in  $T_{p}$, with the result (at zero temperature)
\begin{eqnarray}
I=\frac{e\Delta_{0}}{2\hbar}\sin\phi\sum_{p=1}^{N}T_{p}=
\frac{\pi\Delta_{0}}{2e}G\sin\phi ,
\label{Iphi3}
\end{eqnarray}
which is also plotted for comparison in Fig.\ 2 (dotted curve). The
corresponding critical current is $(\pi\Delta_{0}
/2e)G$. The well-known result (\ref{Iphi3}) for a tunnel junction \cite{Amb63}
differs from the result (\ref{Iav2}) for a disordered weak link
\cite{Kul75}. Za\u{\i}tsev \cite{Zai84} has explained the difference in terms
of different boundary conditions for the Greens functions at the interface
between the junction and the bulk superconductor. The present work yields an
alternative perspective on this issue: Although the tunnel junction and the
disordered weak link may have the same conductance (i.e.\ the same value of
${\rm Tr}\, tt^{\dagger}$), {\em the supercurrents differ because the
distribution of transmission eigenvalues is different}. In a tunnel junction
all $N$ eigenvalues are $\ll 1$, while in the disordered system a fraction
$l/L$ of the eigenvalues is of order unity, the remainder being exponentially
small \cite{Sto91}.

\subsubsection{Supercurrent Fluctuations}

The analysis of Kulik and Omel'yanchuk is based on a
diffusion equation for the {\em ensemble-averaged\/} Greens
function, and can not therefore describe the mesoscopic
fluctuations of $I(\phi)$ from the average. In contrast,
Eq.\ (\ref{Iphi2}) holds for a {\em specific\/}
member of the ensemble of impurity configurations.
The statistical properties of the transmission eigenvalues $T_{n}$
in this ensemble are known from the theory of conductance fluctuations
\cite{Alt91}. A general result \cite{Sto91,Imr86b} is that a {\em linear
statistic\/} ${\cal L}(f)\equiv\sum_{n=1}^{N}f(T_{n})$ on the eigenvalues (with
$f(T)\rightarrow 0$ for $T\rightarrow 0$ and $f(T)={\cal O}(1)$ for $T\ltequiv
1$) fluctuates with root-mean-square value of order unity
\begin{equation}
{\rm rms}\,{\cal L}\equiv\left( \langle{\cal L}^{2}\rangle -\langle{\cal
L}\rangle^{2}\right)^{1/2}={\cal O}(1),\label{rmsL}
\end{equation}
independent of $N$, $L$, or $l$ as long as $l\ll L\ll Nl$. An additional
requirement is that the junction is not much wider than long:
$W_{y},W_{z}\ltequiv L$ where $W_{y}$ and $W_{z}$ are the widths in the $y$-
and $z$-direction. If the junction is much wider
than long in one direction, then ${\rm rms}\,{\cal L}$
is of order $(W_{y}W_{z}/L^{2})^{1/2}$.
It is not necessary for these results that $f(T)$ is itself a linear function
of $T$. In the special case that $f(T)=T$, the coefficient on the
right-hand-side of Eq.\ (\ref{rmsL}) is known \cite{Lee85}:
\begin{equation}
{\rm rms}\,\sum_{n=1}^{N}T_{n}\equiv C_{\rm UCF}=0.365,\;{\rm if}\;
W_{y},W_{z}\ll L.\label{CUCF}
\end{equation}

According to Eq.\ (\ref{Iphi2}) the supercurrent $I(\phi)$ is a linear
statistic on $T_{n}$, so that we conclude that (at zero temperature) ${\rm
rms}\, I(\phi)\simeq e\Delta_{0}/\hbar$ times a function of $\phi$ which is
independent of the length of the junction or the degree of disorder.
At small $\phi$ we may linearize in $T$ and use Eq.\ (\ref{CUCF}), with the
result
\begin{eqnarray}
{\rm rms}\,I(\phi)=
\case{1}{2} C_{\rm UCF}(e\Delta_{0}/\hbar)\phi\tanh
(\Delta_{0}/2k_{\rm B}T)+{\cal O}(\phi^{2}).
\label{rmsIphi}
\end{eqnarray}

The critical current $I_{\rm c}\equiv{\rm max}\, I(\phi)\equiv I(\phi_{\rm c})$
is not by definition a linear statistic on $T_{n}$, since the phase $\phi_{\rm
c}$ at which the maximum supercurrent is reached depends itself on all the
transmission eigenvalues. It is therefore not possible in general to write
$I_{\rm c}$ in the form $\sum_{n}f(T_{n})$, as required for a linear statistic.
However, $I_{\rm c}$ does become a linear statistic in the limit $L/Nl\ll 1$
(or equivalently $\langle G\rangle\gg e^{2}/h$) of a good conductor (which is
the appropriate limit for mesoscopic fluctuations). To see this, we write
\begin{equation}
I(\phi)=\langle I(\phi)\rangle [1+\epsilon X(\phi)],\;\epsilon\equiv
L/Nl,\label{Xphi}
\end{equation}
where the function $X(\phi)$ accounts for the sample-to-sample fluctuations of
$I(\phi)$ around the ensemble average $\langle I(\phi)\rangle$. One has
$\langle X\rangle =0$, ${\rm rms}\, X={\cal O}(1)$. We now expand $I_{\rm c}$
to lowest order in $\epsilon$. Defining $\phi_{\rm c}\equiv\phi_{\rm
c}^{(0)}+\epsilon\phi_{\rm c}^{(1)}$, where ${\rm max}\,\langle
I(\phi)\rangle\equiv\langle I(\phi_{\rm c}^{(0)})\rangle$, one can write
\begin{eqnarray}
I_{\rm c}&\equiv &\left\langle I(\phi_{\rm c})\right\rangle\left[ 1+\epsilon
X(\phi_{\rm c})\right]\nonumber\\
&=&\left\langle I(\phi_{\rm c}^{(0)}+\epsilon\phi_{\rm c}^{(1)})\right\rangle
\left[ 1+\epsilon X(\phi_{\rm c}^{(0)}+\epsilon\phi_{\rm
c}^{(1)})\right]\nonumber\\
&=&\left\langle I(\phi_{\rm c}^{(0)})\right\rangle \left[ 1+\epsilon
X(\phi_{\rm c}^{(0)})+{\cal O}(\epsilon^{2})\right] =I(\phi_{\rm
c}^{(0)}).\label{Icepsilon}
\end{eqnarray}
In the third equality we have used that, by definition, $d\langle I\rangle
/d\phi=0$ at $\phi =\phi_{\rm c}^{(0)}$. Since $I(\phi_{\rm c}^{(0)})$ is a
linear statistic on $T_{n}$ (note that $\phi_{\rm c}^{(0)}$ is by definition
independent of $T_{n}$), we conclude that the critical current $I_{\rm c}$ is a
linear statistic on the transmission eigenvalues in the limit $\epsilon\equiv
L/Nl\rightarrow 0$, and hence that ${\rm rms}\, I_{\rm c}={\rm rms}\,
I(\phi_{\rm c}^{(0)})\simeq e\Delta_{0}/\hbar$. The value of the numerical
coefficient remains to be calculated.

Experimentally, sample-to-sample fluctuations are not
easily studied. Instead, fluctuations in a given sample
as a function of Fermi energy $E_{\rm F}$ are more accessible. Josephson
junctions consisting of a two-dimensional electron gas (2DEG) with
superconducting
contacts allow for variation of $E_{\rm F}$ in the 2DEG by means of
a gate voltage \cite{Tak85}.
Point-contact junctions can be defined in
the 2DEG either lithographically or
electrostatically (using split gates) \cite{Eer91}.
For such a system one would expect that if
$E_{\rm F}$ is varied on the scale
of $E_{\rm c}$, the low-temperature critical current
will fluctuate by an amount
of order $e\Delta_{0}/\hbar$, independent of the properties
of the junction.

\section*{Acknowledgements}
Part of the work reviewed in this paper was performed in collaboration with H.
van Houten at the Philips Research Laboratories in Eindhoven, The Netherlands.
Research at the University of Leiden is supported by the ``Stichting
Fundamenteel Onderzoek der Materie'' (FOM), with financial support from the
``Stichting Nederlands Wetenschappelijk Onderzoek'' (NWO).

\end{document}